\documentclass[3p,times,procedia]{elsarticle}
\usepackage{nupha_ecrc}

\volume{00}

\firstpage{1}

\journalname{Nuclear Physics A}

\runauth{}

\jid{nupha}

\jnltitlelogo{Nuclear Physics A}

\usepackage{amssymb}
\usepackage[figuresright]{rotating}

\begin{document}
\newcommand{\trento}{T$\mathrel{\protect\raisebox{-2.1pt}{R}}$ENTo}

\begin{frontmatter}



%% Instructions from Editor: Please use the following \dochead only in the preprint version (e-print arXiv etc.); 
%% use empty \dochead{} when submitting to Nuclear Physics A!
\dochead{XXVIIIth International Conference on Ultrarelativistic Nucleus-Nucleus Collisions\\ (Quark Matter 2019)}

\title{Effects of initial state fluctuations on the mean transverse momentum}


\author[label2]{Fernando G. Gardim}
\author[label1]{Giuliano Giacalone}
\author[label3]{Matthew Luzum}
\author[label1]{Jean-Yves Ollitrault}

\address[label2]{Instituto de Ci\^encia e Tecnologia, Universidade Federal de Alfenas, 37715-400 Po\c cos de Caldas, MG, Brazil}
\address[label1]{Universit\'e Paris-Saclay, CNRS, CEA, Institut de physique th\'eorique, 91191, Gif-sur-Yvette, France}
\address[label3]{Instituto de F\'{i}sica, Universidade de S\~{a}o Paulo, 
R. do Mat\~{a}o 1371, 05508-090  S\~{a}o Paulo, SP, Brazil}

\begin{abstract}
  We show that in ideal hydrodynamic simulations of heavy-ion collisions, initial state fluctuations result in an increase of the mean transverse momentum of outgoing hadrons, $\langle p_t\rangle$. 
  Specifically,  $\langle p_t\rangle$ is larger by a few percent if realistic fluctuations are implemented than with smooth initial conditions, the multiplicity and impact parameter being kept fixed. 
  We show that result can be traced back to the fact that for a given total entropy, the initial energy contained in the fluid is larger if the density profile is bumpy.  
  We discuss the implication of these results for the extraction of the equation of state of QCD from experimental data. 
\end{abstract}

\begin{keyword}
  heavy ions \sep
  nucleus-nucleus collisions \sep
  LHC \sep
  equation of state
%% keywords here, in the form: keyword \sep keyword2
\end{keyword}
\end{frontmatter}

% \linenumbers


\section{Introduction}
%\label{}

The importance of event-to-event fluctuations of the initial density profile created in realtivistic heavy-ion collisions has been demonstrated in studies of anisotropic flow~\cite{Alver:2006wh,Takahashi:2009na,Alver:2010gr}.
Here, we study their effect on the mean transverse momentum, $\langle p_t\rangle$,  of outgoing hadrons.
We assume for simplicity that the evolution of the system is driven by ideal hydrodynamics~\cite{Kolb:2003dz,Romatschke:2017ejr}. 
It has long been recognized that initial-state fluctuations, followed by ideal hydrodynamic expansion, enhance the production of high-$p_t$ particles~\cite{Andrade:2008xh,Holopainen:2010gz}.
This is understood as the effect of larger pressure gradients in the initial state, leading to higher fluid velocities. 
Here, we do not study the detailed structure of the $p_t$ spectrum, but focus on the mean $p_t$.
One of our goals is to check whether the tight correlation between $\langle p_t\rangle$ and the energy per particle, which has been established with smooth initial conditions, and which can be used to extract the equation of state from experimental data~\cite{VanHove:1982vk,Gardim:2019xjs}, is preserved in the presence of fluctuations.

\section{Event-by-event hydrodynamics}

\begin{figure}[h]
\begin{center}
\includegraphics[width=\linewidth]{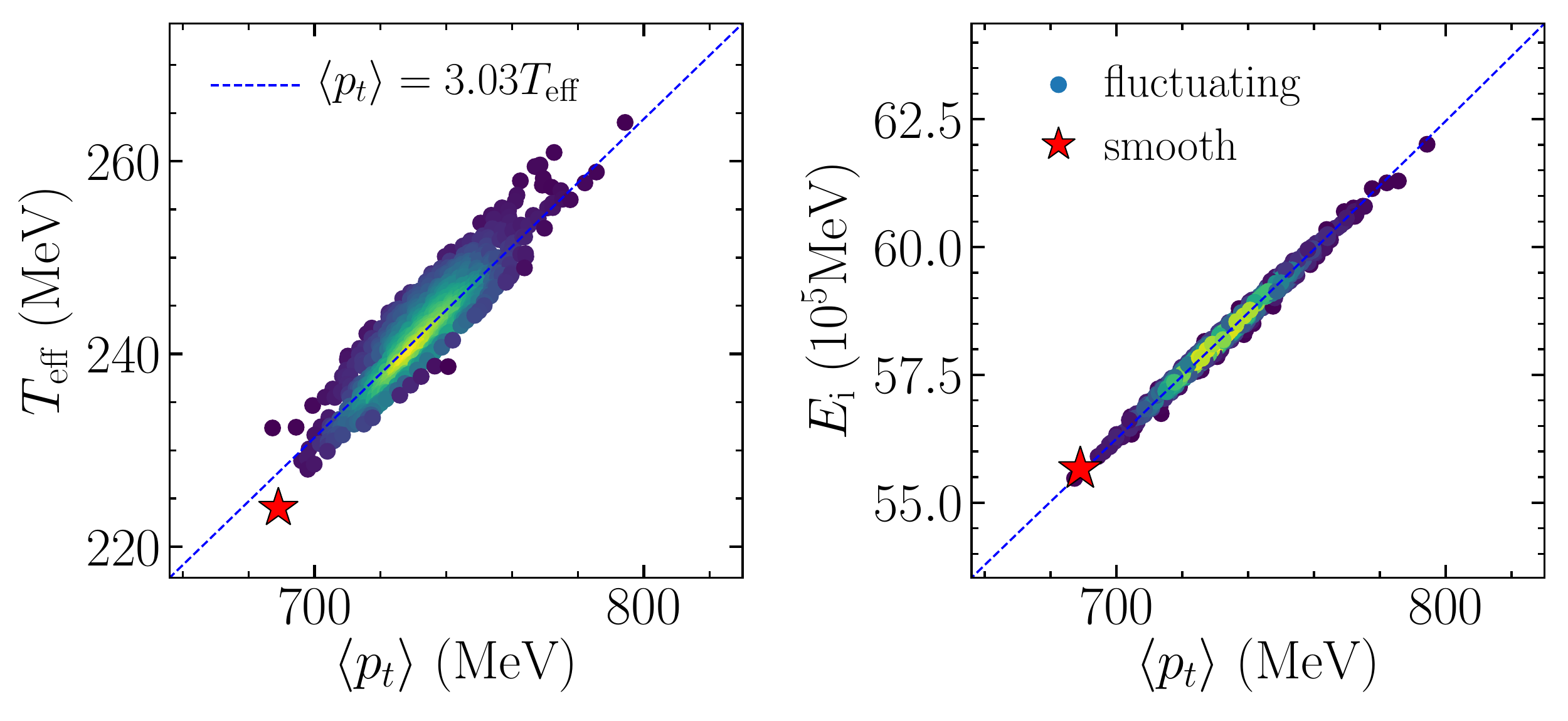} 
\end{center}
\caption{(Color online) 
\label{fig:figure}
Results of inviscid fluid-dynamical calculations for 850 events with the same impact parameter and entropy, but where the distribution of entropy in the transverse plane fluctuates event to event. 
Each filled circle corresponds to a single event. The red stars correspond to the event-averaged (smooth) initial density profile.
Left: Scatter plot of the mean transverse momentum $\langle p_t\rangle$ and the effective temperature $T_{\rm eff}$, defined by Eq.~(\ref{effective}).
Right: Scatter plot of $\langle p_t\rangle$ and initial energy per unit rapidity $E_i$ in our event-by-event hydrodynamic calculations. 
The dotted line in the right panel is a one-parameter fit using $E_i=\alpha \epsilon(T)/s(T)$, where $T\equiv \langle p_t\rangle/3.03$, and $\alpha$ is a fit parameter (see text). 
}
\end{figure}

We model event-to-event fluctuations using the \trento{} parameterization of the  initial entropy density profile~\cite{Moreland:2014oya}, where parameters are tuned to LHC data~\cite{Giacalone:2017uqx}.
We generate 850 events with the same impact parameter ($b=2.5$~fm) and the same total entropy, corresponding to the mean entropy of Pb+Pb collisions at $\sqrt{s_{NN}}=5.02$~TeV in the 0-5\% centrality window. 
Since entropy is conserved in ideal hydrodynamics, and the final multiplicity is proportional to the entropy to a very good approximation, this in turn implies that the number of particles is almost identical in every event (we have checked that the rms deviation of the multiplicity is smaller than 1\%).
For each event, the initial entropy density profile is evolved using the MUSIC~\cite{Paquet:2015lta} hydrodynamic code, assuming longitudinal boost invariance~\cite{Bjorken:1982qr}, and an initial time $\tau_0=0.6$~fm/c.
The fluid freezes out~\cite{Cooper:1974mv} into individual hadrons at the temperature $T_f=156.5$~MeV~\cite{Bazavov:2018mes}. Resonance decays are implemented, and $\langle p_t\rangle$ is evaluated for charged particles after decays.
%am I forgetting something, like, pt or eta cuts?
We also run, for the sake of comparison, a single event with the same impact parameter and entropy, but a smooth entropy density profile corresponding approximately to the average over all events of the fluctuating profiles. 
Figure~\ref{fig:figure} shows that $\langle p_t\rangle$ is larger with fluctuations (circles) than with smooth initial conditions (star), on average by $7\%$.

\section{Interpretation}
The mean transverse momentum is closely related to the energy per particle~\cite{Gardim:2019xjs}. 
Since all our events have the same multiplicity by construction, this  suggests that $\langle p_t\rangle$ is determined by the fluid energy at freeze-out, $E_f$, which fluctuates event to event. 
To study the correlation between $\langle p_t\rangle$ and $E_f$, we follow Ref.~\cite{Gardim:2019xjs}. We define the {\it effective\/} temperature, $T_{\rm eff}$, and the effective volume, $V_{\rm eff}$, as those of a uniform fluid at rest which would have the same energy and entropy as the fluid at freeze-out: 
\begin{eqnarray}
  \label{effective}
E_f=\int_{\rm f.o.} T^{0\mu} d\sigma_\mu&=&\epsilon(T_{\rm eff}) V_{\rm eff},\cr
S_f=\int_{\rm f.o.} s u^{\mu} d\sigma_\mu&=&s(T_{\rm eff}) V_{\rm eff}.
\end{eqnarray}
For fixed entropy $S_f$, the energy $E_f$ is to a good approximation proportional to $T_{\rm eff}$, so that fluctuations of $E_f$ correspond to fluctuations of $T_{\rm eff}$. 
Figure~\ref{fig:figure} (left) shows that $\langle p_t\rangle$ is tightly correlated with $T_{\rm eff}$, and that the relation $\langle p_t\rangle\simeq 3.03~T_{\rm eff}$ is a good approximation for all events, i=cluding the smooth event. 
Therefore, the larger $\langle p_t\rangle$ in the presence of fluctuations can be attributed to a larger $T_{\rm eff}$ or, equivalently, a larger $E_f$. 

Now, the reason why fluctuations increase the final energy $E_f$ is simply that they also increase the initial energy $E_i$, evaluated by integrating the energy density over the fluid volume at time $\tau_0$.  
Figure~\ref{fig:figure} (right) displays the scatter plot of $\langle p_t\rangle$  and $E_i$ for our sample of events. 
Comparison between the two panels shows that the correlation of $\langle p_t\rangle$ with $E_i$ is even stronger than the correlation with  $T_{\rm eff}$. 
More specifically, if one replaces $E_f=\int_{\rm f.o.} T^{0\mu} d\sigma_\mu$ with $E_f=xE_i$ in  Eq.~(\ref{effective}) (where $x\simeq 0.41$ is the average fraction of energy remaining after longitudinal cooling), and if one denotes by $T'_{\rm eff}$ the new value of $T_{\rm eff}$  obtained by solving the equations, then $3.03T'_{\rm eff}$ is a better approximation of $\langle p_t\rangle$, represented by the dotted line in Fig.~\ref{fig:figure} (right), than $3.03T_{\rm eff}$ in Fig.~\ref{fig:figure} (left).
It is a paradox that the initial energy is a better predictor of $\langle p_t\rangle$ than the energy at freeze-out, since particle emission takes place at freeze-out. 
We do not have a simple explanation for this striking observation. 

Fluctuations increase the initial energy $E_i$ for the following reason. 
At a given point in the transverse plane, the entropy density $s$ fluctuates event to event. 
The energy density at this point, denoted by $\epsilon$,  is a function of $s$.
Thermodynamic stability requires that the function $\epsilon(s)$ is convex. 
This implies that the average value of $\epsilon$ over events is larger than the value of $\epsilon$ corresponding to the average value of $s$, that is, larger than the energy density of the smooth profile. 
The relative increase depends on the model of fluctuations,  in particular the transverse size of the inhomogeneities. 
A more spiky profile will result in a larger increase. 
Note that in contrast, anisotropic flow is typically insensitive to the scale of inhomogeneities~\cite{Gardim:2017ruc}.
The relative increase also depends on how thermalization is achieved in the early stages of the collisions. 
Assuming that ideal hydrodynamics holds at $\tau_0=0.6$~fm/c, as we do in our calculation, is likely to overestimate the 
energy increase. 
In order to obtain a realistic estimate of this effect, it is crucial to carefully model the thermalization phase~\cite{Kurkela:2018wud}.

\begin{figure}[t]
\begin{center}
\includegraphics[width=.7\linewidth]{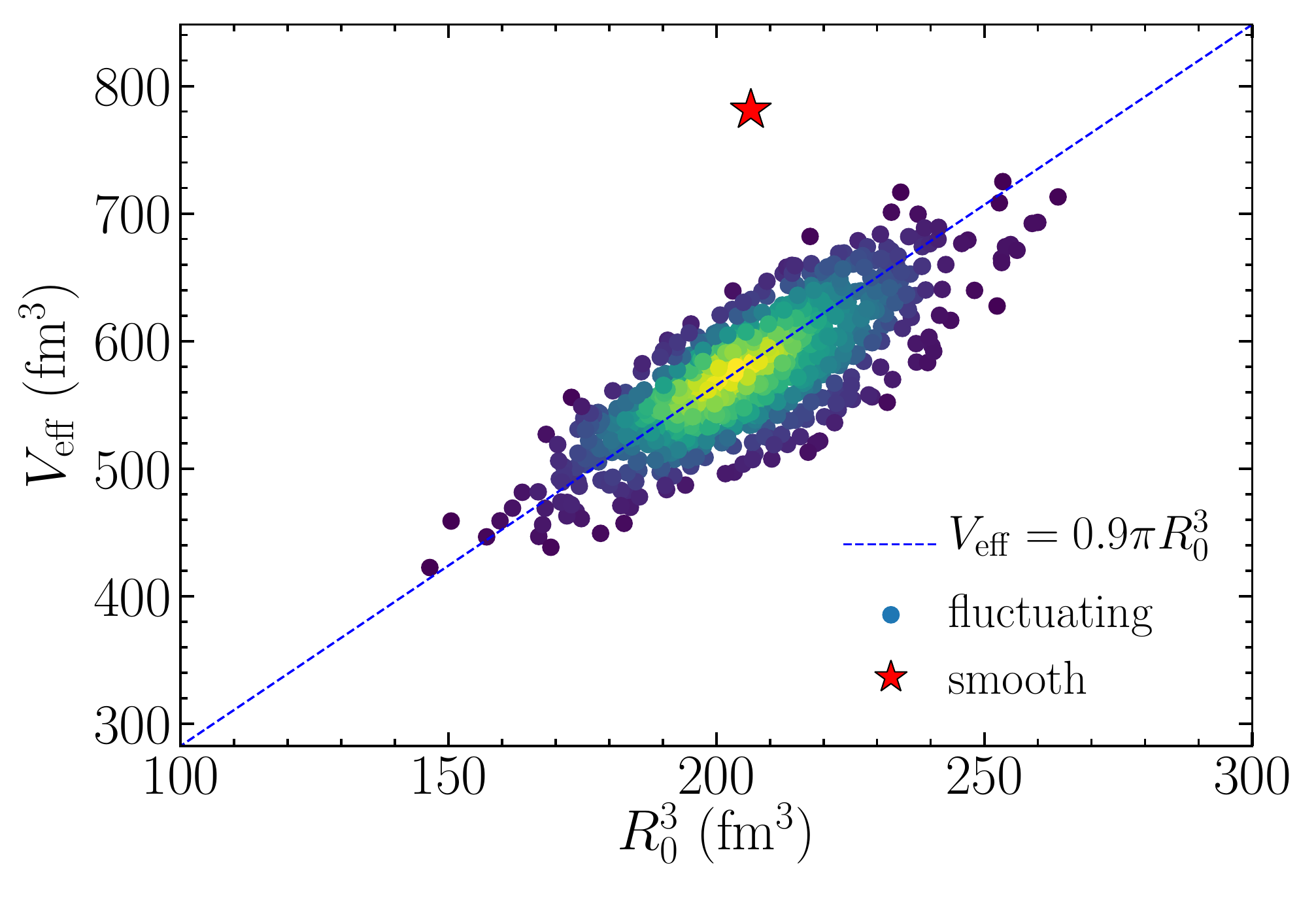} 
\end{center}
\caption{(Color online) 
\label{fig:figure2}
 Scatter plot of $R_0^3$ and $V_{\rm eff}$, where $R_0$ is the initial rms radius defined by Eq.~(\ref{defR0}), and $V_{\rm eff}$ is the effective volume of the quark-gluon plasma per unit rapidity, as defined by Eq.~(\ref{effective}). As in Fig.~\ref{fig:figure}, each filled circle corresponds to a single event, and the red star corresponds to the event-averaged initial density profile.   
 }
\end{figure}

\section{Effective volume}
Since all events have the same total entropy $S_f$, inspection of Eq.~(\ref{effective}) shows that larger values of the effective temperature $T_{\rm eff}$ imply smaller values of the effective volume $V_{\rm eff}$.
This is confirmed by the numerical results displayed in Fig.~\ref{fig:figure2}.
Fluctuating initial conditions result in a smaller effective volume than smooth initial conditions.
In Ref.~\cite{Gardim:2019xjs}, it was argued that the effective volume is determined by the initial radius $R_0$, defined by:
\begin{equation}
  \label{defR0}
(R_0)^2\equiv \frac{2\int_{\bf r} |{\bf r}|^2 s(\tau_0,{\bf r})}{\int_{\bf r} s(\tau_0,{\bf r})},
\end{equation}
where $s(\tau_0,{\bf r})$ is the initial entropy density, and the integration runs over the transverse plane.
%The factor $2$ in the numerator ensures that for a uniform density profile in a circle of radius $R_0$, the right-hand side gives $(R_0)^2$.
The results  in Fig.~\ref{fig:figure2} show that the value of $R_0^3$ for smooth initial conditions, indicated by a star, corresponds to the average of $R_0^3$ for fluctuating initial conditions, as expected. 
On the other hand, the proportionality factor $V_{\rm eff}/R_0^3$ is consistently smaller if fluctuations are present. 
This is a corollary of the increase of the initial energy discussed above: 
Thus, the magnitude of the effect depends on the transverse scale of the fluctuations. 

\section{Implications for the extraction of the equation of state}
Figure~\ref{fig:figure} (left) shows that the proportionality factor between $T_{\rm eff}$  
and $\langle p_t\rangle$ is essentially the same with smooth or fluctuating initial conditions. 
Thus, the value of $T_{\rm eff}=222\pm 9$~MeV obtained in Ref.~\cite{Gardim:2019xjs} assuming smooth initial conditions is robust with respect to the inclusion of initial state fluctuations. 
In order to estimate the corresponding  entropy density $s(T_{\rm eff})$ from the second line of Eq.~(\ref{effective}), however, one needs the value of $V_{\rm eff}$, which cannot be measured, and must be calculated. 
The present study suggests that the effect of initial fluctuations on $V_{\rm eff}$ is potentially large. 
It is likely to be overestimated by the present study, where we use ideal hydrodynamics at early times, while a large pressure anisotropy is expected due to the fast longitudinal expansion. 
A quantitative study will require to include transport coefficients and  a dynamical treatment of the thermalization phase~\cite{Kurkela:2018wud}.

\section*{Acknowledgments}
F.G.G. was supported by Conselho Nacional de Desenvolvimento Cient\'{\i}fico  e  Tecnol\'ogico  (CNPq grant 205369/2018-9 and 312932/2018-9). 
M.L.~acknowledges support from FAPESP projects 2016/24029-6  and 2017/05685-2.
F.G.G. and  M.L.  acknowledge support from project INCT-FNA Proc.~No.~464898/2014-5 and
G.G., M.L. and J.-Y.O.   from  USP-COFECUB (grant Uc Ph 160-16, 2015/13).


\bibliographystyle{elsarticle-num}
\bibliography{qm.bib}






% \end{thebibliography}

\end{document}